\documentclass[useAMS,usenatbib,usegraphicx]{mn2e}

\usepackage{epsfig}
\usepackage{myaasmacros}

\title[iVINE - Ionization in the tree/SPH code VINE]{iVINE -
  Ionization in the parallel tree/SPH code VINE: First results on the observed age-spread around O-stars}
\author[M. Gritschneder, T. Naab, A. Burkert, S. Walch, F. Heitsch, M. Wetzstein]{M. Gritschneder$^{1}$\thanks{E-mail:
gritschm@usm.uni-muenchen.de}, T. Naab$^{1}$, A. Burkert$^{1}$,
  S. Walch$^{1}$,  F. Heitsch$^{2}$, M. Wetzstein$^{3}$ \\
$^{1}$Universit\"ats-Sternwarte M\"unchen, Scheinerstr.1, D-81679
  M\"unchen, Germany \\ $^{2}$Department of Astronomy,
  University of Michigan, Michigan, United States \\  $^{3}$Department
  of Astrophysical Sciences, Princeton University, Princeton, United States}

\begin{document}

\date{Accepted ???. Received ??? in original form ???}

\pagerange{\pageref{firstpage}--\pageref{lastpage}} \pubyear{2008}

\maketitle

\label{firstpage}

\begin{abstract}
We present a three-dimensional, fully parallelized, efficient
implementation of ionizing UV radiation for smoothed particle
hydrodynamics (SPH) including self-gravity. Our method is based on the
SPH/tree code VINE. We therefore call it iVINE (for Ionization + VINE). This
approach allows detailed high-resolution studies of
the effects of ionizing radiation from e.g. young massive stars on 
their turbulent parental molecular clouds. In this paper 
we describe the concept and the numerical implementation of the
radiative transfer for a plane-parallel geometry and we discuss several
test cases demonstrating the efficiency and accuracy of the new
method. As a first application, we study the radiatively driven
implosion of marginally stable molecular clouds at various distances 
of a strong UV source and show that they are driven into gravitational
collapse. The resulting cores are very compact and dense exactly as it
is observed in clustered environments.
Our simulations indicate that the time of triggered collapse depends on the
distance of the core from the UV source. Clouds closer to the source
collapse several $10^5$ years earlier than more distant clouds. This effect can
explain the observed age spread in OB associations where stars closer to the
source are found to be younger. 
We discuss possible uncertainties in the observational derivation of
shock front velocities due to early stripping of proto-stellar
envelopes by ionizing radiation.
\end{abstract}


\section{Introduction}
\label{intro}
As hydrodynamical simulations
become more and more advanced one of the key issues is the successful
implementation of additional physics like the effects of
radiation. Prominent applications are for example the reionization of
the early universe (for a comparison of methods see
\citealt{2006MNRAS.371.1057I} and references therein).

In our present day universe ionizing radiation still plays a vital
role. UV-radiation from massive, young stars ionizes their
surrounding. The hot, ionized gas then expands into the cold, neutral
gas and thus drives shock fronts into the parental molecular
clouds.  Up to now it is not fully understood if this violent
feedback enhances or hinders star
formation. \citet{1977ApJ...214..725E} proposed that the shock front
builds up dense regions by sweeping up the cold gas, which then
eventually collapse  
due to gravitational instability and form stars. This is called the
'collect and collapse model' (see also the review by
\citealt{1998ASPC..148..150E}). Another  
situation arises when preexisting, dense structures (e.g. molecular cloud
cores) that are gravitationally marginally stable get compressed by the
approaching front and start collapsing. This is commonly called radiation
driven implosion (see e.g. \citealt{1982ApJ...260..183S}).

Observations provide widespread evidence for these scenarios
(\citealt{1989ApJ...342L..87S}, \citealt{1996AJ....111.2349H}). More
recent observations indicate
triggered star formation on the edges of HII-regions e.g. in the
Orion clouds \citep{2002A&A...393..251S}, the Carina nebula
\citep{2000ApJ...532L.145S}, M16 \citep{2002ApJ...568L.127F}, M17
\citep{2002ApJ...577..245J}, 30 Dor \citep{2002AJ....124.1601W} and
the SMC \citep{2007arXiv0710.1352G}. \citet{2005A&A...433..565D}
report triggered star formation in samples of more distant HII 
regions. Besides these quite complex large scale regions there have
been numerous observations of bright rimmed cometary globules. These
are small isolated clouds with a clear head to tail 
structure with the dense heads pointing towards an ionizing
source \citep{1991ApJS...77...59S}. Their morphology enables a
direct comparison to simulations. In particular, the properties
of individual young stellar objects (YSOs) surrounding OB-associations
can be determined precisely. 
YSOs are observed in the mass range from T Tauri
($0.1-3 \textrm{M}_{\odot}$) up to Herbig Ae Be ($2-8\textrm{
  M}_{\odot}$) stars (see e.g. \citealt{2007ApJ...657..884L},
\citealt{2007arXiv0711.1515S}). The velocity of the shock front
triggering the star formation is calculated from the age difference of
the stars and their relative distance. These estimates are in the
range of a few km/s (e.g. \citealt{2004A&A...414.1017T},
\citealt{2007ApJ...654..316G}).

Numerous simulations on the topic of cloud evaporation and sequential star
formation have been performed. \cite{1982A&A...108...25Y}
(and references therein) published a series of two-dimensional
simulations on the gas dynamics of HII-regions, especially on champagne
flows, where a stream of hot gas breaks through the border of cold,
confining gas. Subsequently, \citet{1995ApJ...451..675E}  
presented two-dimensional, grid-based simulations showing that
the expansion of an HII-region into the surrounding cloud can trigger star
formation. 
\citet{2003MNRAS.338..545K} demonstrated with a three-dimensional SPH code, 
that a marginally stable molecular cloud core can be triggered into collapse
when exposed to strong UV radiation.
With a more detailed description of radiation implemented into an
SPH code \cite{2006MNRAS.369..143M} 
could reproduce the observed features of the
Eagle Nebula, including the photodissociation regions and the
temperature profile.
Using a three-dimensional grid-based scheme,
\citet{2006ApJ...647..397M} simulated 
the HII-region excavated by a point source of UV-radiation.  They find remarkably similar morphologies
and physical properties when comparing their models to observations. 
Furthermore simulations with an SPH-code by \citet{2005MNRAS.358..291D}
and a grid code by \citet{2007ApJ...668..980M} showed that a turbulent
interstellar medium surrounding an O-star allows the ionizing
radiation to efficiently expel most of the nearby gas. Only the 
denser regions resist and continue to collapse. 

However none of the authors so far described ionizing radiation as an
efficient trigger for star formation.
There is only weak evidence by \citet{2007MNRAS.377..535D},
that the external irradiation of a collapsing cloud by a point
source can indeed increase the star formation efficiency from 3\% to 4\%
when compared to a control run without radiation. For a review of 
feedback processes we refer the reader to \citet{2007arXiv0711.4047M}.
For completeness we would like to refer to recently published
 implementations for ionizing radiation into an SPH code 
by \citet{2008MNRAS.389..651P}, where the photons of a source are
followed along cones, and 
\citet{2008MNRAS.386.1931A}, where the radiation is followed via a
Monte Carlo ray-tracing scheme

All these studies demonstrate that there is a strong connection between
the UV-radiative feedback from massive stars and the observed morphologies of the 
ambient molecular cloud gas. 
Yet, a quantitative discussion of the interaction between UV-radiation
and turbulent molecular clouds is still missing. To advance our understanding,
we introduce 
iVINE, the fully parallel implementation of UV-radiation in the
parallel tree-SPH-code VINE (\citealt{2008arXiv0802.4245W},
\citealt{2008arXiv0802.4253N}). This efficient tool permits high
resolution simulations of molecular clouds 
in the vicinity of strong UV sources such as an O-star or association. 

The paper is structured as follows. The physical model and its
implementation are described in Section \ref{NM},  
followed by a detailed comparison of the scheme with analytical results (Sec.~\ref{NT}).
We apply the new method to the radiatively driven implosion of a marginally stable
molecular cloud core and compare three simulations with different
initial UV fluxes to observations 
(Sec~\ref{simulations}). In Section~\ref{summary} we summarize and
discuss the results.

\section{Numerical Method}
\label{NM}
As soon as a young massive star emits UV-radiation it ionizes its
surrounding, creating a HII-region. Initially
the ionization proceeds fast with a speed of this rarefied (or
R-type) front of $v_\mathrm{R}>>a_\mathrm{hot}$, where
$a_\mathrm{hot}$ is the sound speed of the hot, ionized gas. 
After a sound crossing timescale the hot gas
reacts to its increased temperature and an isothermal shock front is
driven into the cold surrounding medium. This dense (or D-type) shock
travels at a much smaller speed $v_\mathrm{D} \approx
a_\mathrm{hot}$. For a full textbook analysis of this evolution see
e.g. \citet{1989agna.book.....O}.

\subsection{Prescription of ionizing radiation}
\label{basics}
To follow the evolution of the HII-region of a young
massive star in a numerical simulation we use a prescription for the
ionizing UV-radiation 
similar to the one proposed by \citet{2000MNRAS.315..713K} as
presented before \citep{2007IAUS..237..246G}. The flux
$J$ at any given position $x$ is given by
\begin{equation}
\label{eq_tau}
J(x)=J_\mathrm{Ly}e^{-\tau_{\nu}(x)},
\end{equation}
where $J_\mathrm{Ly}$ is the Lyman continuum flux of the hot star. The optical
depth $\tau_{\nu}$ is given by the integral along the line of sight
between the source of radiation and the position $x$
\begin{equation}
\label{eq_def_tau}
\tau_{\nu}=\int_0^{x}\kappa_{\nu}\rho dx,
\end{equation}
where $\kappa_{\nu}$ is the frequency weighted absorption coefficient
\begin{equation}
\kappa_{\nu} = \frac{\sigma_{\nu}n_\mathrm{H}}{\rho},
\end{equation}
with $n_\mathrm{H}$ being the number density of neutral hydrogen and $\rho$ the
mass density of the gas. We assume the gas is pure hydrogen with a
mean molecular weight of $\mu =1$.
As the frequency dependent absorption cross
section $\sigma_{\nu}$ peaks at the
Lyman break it is a valid assumption to take an average cross section
$\bar{\sigma}$, thereby approximating the radiation to be
monochromatic. Thus, every photon above the Lyman break is assumed to
ionize one hydrogen atom. 

We define the ionization degree $\eta$ as
\begin{equation}
\label{def_eta}
\eta = \frac{n_\mathrm{e}}{n},
\end{equation}
where $n_\mathrm{e}$ is the number density of electrons and $n$ is the combined
number density of protons and neutral hydrogen atoms. The time
derivative of the ionization degree can be written as
\begin{equation}
\label{eq_delta_dt}
\frac{\mathrm{d}\eta}{\mathrm{d}t} =
\frac{1}{n}\frac{\mathrm{d}n_\mathrm{e}}{\mathrm{d}t} = \frac{1}{n}(I-R),
\end{equation}
with the ionization rate $I$ given as
\begin{equation}
I = \nabla J
\end{equation}
and the recombination rate $R$ as
\begin{equation}
\label{eq_rec}
R = n_\mathrm{e}^2 \alpha_\mathrm{B} = \eta^2 n^2 \alpha_\mathrm{B}.
\end{equation}
For the recombination coefficient $\alpha_\mathrm{B}$ we choose
\begin{equation}
\alpha_\mathrm{B}  = \sum_{i=2}^{\infty} \alpha_i,
\end{equation}
where $\alpha_i$ is the recombination probability for a level i of the
hydrogen atom.
The recombination of electrons and protons leads to a diffuse field
of Lyman continuum photons, which in turn can again ionize a hydrogen
atom. We neglect this effect under the assumption that every reemitted
photon is in turn immediately absorbed in the direct surrounding. This
assumption, called `on the spot approximation' is 
valid as long as the hydrogen density is high enough
(e.g. \citealt{1978ppim.book.....S}), which is always
true in the vicinity of the ionization front.
Some fraction of the UV-photons is absorbed by dust, and re-emitted in
the IR-regime, leading to an effective lower flux. We neglect this
effect, since the flux incident on the simulation volume is determined
largely by  its distance from the radiation source, so that
geometrical dilution of the radiation field is likely to be more
important than absorption by dust.

The average temperature of the gas is coupled linearly to the ionization
degree $\eta$ through
\begin{equation}
\label{eq_temperature}
T = T_\mathrm{hot} \cdot \eta + T_\mathrm{cold} \cdot (1-\eta),
\end{equation}
where $ T_\mathrm{cold}$ is the initial temperature of the cold, unionized
gas and  $ T_\mathrm{hot}$ is the average temperature of the ionized gas.

\subsection{Implementation}
\label{implementation}
To treat the hydrodynamical and gravitational evolution of the gas we
use the parallel smoothed particle hydrodynamics (SPH) code called
VINE developed by
\citet{2008arXiv0802.4245W} and \citet{2008arXiv0802.4253N}. SPH is a
Lagrangian  method, which renders it
extremely suitable to cover several orders of magnitude in density and
spatial scale, for
example during cloud core collapse and star formation. VINE is a powerful
parallel implementation of the SPH method in combination with a
tree code for the calculation of gravitational forces. It offers a
Runge-Kutta integrator as well as a 
Leapfrog integrator. Both schemes can be used in combination with
individual particle time-steps. For this work the Leapfrog integrator
is chosen. Every time the equation of state is computed we calculate
the ionization degree for all particles in the entire simulation. 

The heating by
UV-radiation can be treated as decoupled from the dynamic evolution since
the recombination timescale
\begin{equation}
\label{t_rec} t_\mathrm{rec} = \frac{1}{n \alpha_\mathrm{B}}
\end{equation}
is much shorter than any hydrodynamical timescale. In our simulations
(see Section \ref{simulations}) the crossing time even in the hot gas
is $t_\mathrm{hot} \approx 70$kyr, whereas the recombination timescale is
$t_\mathrm{rec} \approx 1$kyr for a number density of
$n=100\textrm{cm}^{-3}$. Thus, it is valid to treat ionization and
hydrodynamics as two separate processes. In other words 
the ionization can be assumed to happen instantaneously.
Frequently updating the ionization degree together with a modified
time-step criterion  (see Section \ref{timestepping}) ensures 
that the radiation is treated correctly on all scales.

To include the effect of UV-radiation we assume a plane-parallel 
geometry, i.e. parallel rays. 
This is valid as long as the distance from the source of radiation
is larger than the dimensions of the area of infall. 
In our simulations the radiation is impinging from the left hand side,
that is from the negative x-direction. To couple ionization to hydrodynamics
we use a ray-shooting algorithm. As the ionizing flux is propagated 
along
the x-direction, we ensure the conservation of flux by dividing the 
$(y,z)$-plane into sub-domains of equal size,
whose extent along the $x$-direction spans the whole simulation domain.
Along each of these sub-domains or rays the flux is transported in a
conservative manner.
To convert the SPH-particle density $\rho_\mathrm{part}$ correctly
into a gas density distribution within these three-dimensional rays  
the volume and thus the diameter $d_\mathrm{part}$ that each
SPH-particle occupies is calculated via the mass of each 
particle $m_\mathrm{part}$:
\begin{equation}
d_\mathrm{part} = 2\cdot\left(\frac{3}{4\pi}\frac{m_\mathrm{part}}{\rho_\mathrm{part}}\right)^{1/3}.
\end{equation}
The width $\Delta y$ and the height $\Delta z$ of the rays or bins is then set
to the average value $\overline{d_\mathrm{part}}$ 
of the particles closest to the source. To determine this value the
first two particles in each ray at the previous time-step are taken
into account. Since this is the region with the lowest density throughout
the entire simulation this guarantees that the bin-size is always larger
than the characteristic particle resolution.
As soon as the ray approaches a
density increase the local $d_\mathrm{part}$ becomes smaller than
$\Delta y$ and $\Delta z$. 
For
\begin{equation}
\label{ref_crit}
 \frac{d_\mathrm{part}}{\Delta y}= \frac{d_\mathrm{part}}{\Delta z} < \frac{1}{2}
\end{equation}
we refine the ray subsequently into four sub-rays to treat the
ionization of high density regions correctly. Each of the sub-rays
inherits the optical depth of the main ray. Currently the code allows
for five levels of refinement, thus increasing the effective bin
resolution in each direction by a factor of 32. In principle it would
be possible to de-refine the sub-rays by using the average optical
depth of the four refined sub-rays for the de-refined bin. We do not
include this, since it would lead to an unphysical shading of lower
density sub-rays as soon as they are combined with a high density
sub-ray due to an overestimation of the optical depth.

To calculate the optical depth, we sort the particles within each bin
according to their distance to the source 
and discretize into subsections of the size
\begin{equation}
\Delta x_{i} = \frac{x_{i+1}-x_{i-1}}{2}.
\end{equation}
Thus, $\Delta x_{i}$ is the projected distance of a particle $i$ to
its direct neighbours closer and further away from the source,
i.e. the length along the line of sight the particle occupies. We then 
calculate the optical depth $\tau$ along each ray 
by summing up the individual optical depths $\tau_i$ of each 
particle $i$. The discrete value of $\tau_i$ is given according to
Eq. \ref{eq_def_tau} as
\begin{equation}
\tau_{i} = \bar{\sigma}\,n_{\mathrm{H},i}\,\Delta x_{i} .
\end{equation}
The number density $n_\mathrm{H}$
and the density $\rho$ used to calculate the recombination 
rate (cf Eq. \ref{eq_rec}) are simply given by
the SPH-density $\rho_\mathrm{part}$. From these quantities the new ionization
degree $\eta$ is determined according to Eq. \ref{eq_def_tau} by  
a Newton-Raphson iteration scheme. It converges with
a precision of more than $0.1\%$ in less than $200$ iterations. 
When the ionization degree reaches a value of $\eta=1\times
10^{-10}$ we terminate the further calculation of this bin.
This implementation is fully parallelized in OpenMP.

\subsection{Modification of the time-step criterion}
\label{timestepping}
A detailed discussion of the different time-step criteria implemented in
the underlying VINE code is given in \cite{2008arXiv0802.4245W}. Note
that our implementation of ionizing radiation is designed to be used
in connection with individual particle time-steps \citep[see][for
  details]{2008arXiv0802.4245W}.
To exactly follow the evolution of a particle during its ionization
process it is vital to use a small enough time-step. To do so we
decided to force every particle to a smaller time-step as soon as
its ionization degree reaches $\eta > 10^{-3}$, i.e. when the particle
is going to be ionized. The new time-step is chosen by a modified
Courant-Friedrichs-Lewy (CFL) condition according to 
\begin{equation}
\Delta t_\mathrm{new}=a_\mathrm{cold}/a_\mathrm{hot}*\Delta t_\mathrm{CFL},
\end{equation}
where $a_\mathrm{cold}$ and $a_\mathrm{hot}$ are the fixed respective
 sound-speeds  of the cold and the ionized gas at $T_\mathrm{cold}$ and
 $T_\mathrm{hot}$. 
$\Delta t_\mathrm{CFL}$ is the individual time-step the particle would
get assigned due to the CFL-condition \citep[see][]{2008arXiv0802.4245W}.
This ensures that the hydrodynamical quantities are treated
correctly even though the particle gets a boost in temperature.
Therefore, we anticipate the subsequent acceleration  due to the
approaching ionization front by choosing already the much smaller
 time-step even though the particle is just ionized to $0.1\%$. 
This criterion also ensures that the ionization degree is followed
 accurately during the evolution of the later dense or D-type
 ionization front, because $v_\mathrm{D}$ is always smaller 
than the sound speed of the hot gas $v_\mathrm{D} <
 a_\mathrm{hot}$. Hence, this front will always be resolved by
 particles which have a small enough time-step to track the hot gas.
In the beginning the evolution of the faster R-type front
 ($v_\mathrm{R} >> a_\mathrm{hot}$) can be followed by using a small enough 
initial time-step since this phase is quite short ($\approx 5$kyr).

The choice of a small initial time-step together with the modified
CFL-criterion ensure that the  ionization degree $\eta$ of a particle 
never changes by more than $\pm0.1$ per time-step in all of our
simulations. Thus, the ionization front can be followed in both
stages (R- and D-type) precisely.

\section{Numerical Tests}
\label{NT} In order to validate the algorithm we
perform  several tests. 
The first series of simulations addresses the evolution of the
Str{\"o}mgren solution and tests whether the time-dependent UV-flux is
treated correctly on all scales. In addition, we demonstrate the correct
implementation of the refinement (Section \ref{tests_no_hydro}). The
second series of simulations (Section \ref{tests_hydro}) is designed to 
demonstrate the correct interaction of ionizing
radiation and hydrodynamics. In the end the successful parallelization
of the code is shown (Section \ref{perform}).
\subsection{Ionization without hydrodynamics}
\label{tests_no_hydro}
\subsubsection{The Str{\"o}mgren test - Ionization by a constant
  UV flux}
\label{stroem}
When hydrodynamics is not taken into account, the homogeneous
surrounding of an ionizing source will always converge towards an
equilibrium between ionization and recombination. 
The volume of the ionized Str{\"o}mgren sphere
\citep{1939ApJ....89..526S} around 
an O-star is given by
\begin{equation}
V_\mathrm{S}=\frac{J_\mathrm{Ly}}{\alpha_\mathrm{B}n^2},
\end{equation}
assuming a monochromatic
source with a constant UV-flux $J_\mathrm{Ly}$ given in photons per
second. $\alpha_\mathrm{B}$ and $n$ are again the recombination coefficient
and the number density (for a textbook analysis see e.g.
\citealt{1991pagd.book.....S}).

In the case of plane-parallel radiation, as discussed
here, this volume is characterized by  the length $x_\mathrm{s}$, which can be
penetrated by the ionizing radiation. $x_\mathrm{s}$ is determined by the surface $S$ on
which the photon flux per area and time, $F_\mathrm{Ly}$, is impinging:
\begin{equation}
\label{x_stroem}
x_\mathrm{s} = \frac{V_\mathrm{s}}{S} =\frac{F_\mathrm{Ly}}{\alpha_\mathrm{B}n^2}.
\end{equation}
The time evolution of the length $x_\mathrm{I}(t)$ of this region is given by the
differential equation 
\begin{equation}
\frac{dx_\mathrm{I}}{dt}n=F_\mathrm{Ly}-x_\mathrm{I}(t)\alpha_\mathrm{B}n^2
\end{equation}
with the solution
\begin{equation}
\label{eq_x_vs_t}
x_\mathrm{I}(t)=x_\mathrm{s}(1-e^{-t/t_\mathrm{rec}}),
\end{equation}
where $t_\mathrm{rec}=1/(n\alpha_\mathrm{B})$ is the recombination or
Str{\"o}mgren timescale.
The shape of the front is given by the ionization equilibrium
equation
\begin{equation}
n(1-\eta)\int_{\nu_\mathrm{Ly}}^{\infty}F_{\nu}\sigma_{\nu}d\nu=n^2\eta^2\alpha_B,
\end{equation}
which can be rewritten for the plane-parallel, monochromatic case in
terms of the ionization degree (cf Eq. \ref{def_eta}) as
\begin{equation}
\label{eq_eta_vs_x}
\frac{d\eta}{dx}=\eta^2\frac{1-\eta}{1+\eta}n\bar{\sigma}x_\mathrm{s}.
\end{equation}
This equation can be solved numerically and gives the ionization degree
$\eta$ at any given position $x$ for the chosen number density $n$
and mean cross-section $\bar{\sigma}$.

To test the code, we ran three simulations: 
\begin{itemize}
\item Case A: $125$k particles placed on a Cartesian grid
\item Case B: $100$k particles placed randomly
\item Case C: $250$k particles placed randomly
\end{itemize}
For cases B and C the particles are
placed randomly in the simulation box and then are
allowed to relax with periodic boundaries and the inclusion of
hydrodynamics for one crossing timescale to
dampen the numeric random noise. Thereafter we switch off the
hydrodynamics and compute the ionization. 
For all simulations we used a mean density
$n = 10\textrm{cm}^{-3}$. 
The simulated volume is $(2\textrm{pc})^3$, the length the ionization can
penetrate is set to $x_\mathrm{s}=1$pc. The recombination coefficient and the
absorption cross-section are set to typical values of
$\alpha_\mathrm{B}=2.7\times10^{-13}\textrm{cm}^3\textrm{s}^{-1}$ and
$\bar{\sigma}=3.52\times10^{-18}\textrm{cm}^2$.
For the above parameters, 
$F_\mathrm{Ly}=8.33\times10^7 \textrm{photons cm}^{-2}\textrm{s}^{-1}$ and
$t_\mathrm{rec}=11.8$kyr.
The simulations run up to 
$t=5t_\mathrm{rec}$ to allow for a quasi-equilibrium state to evolve.
\begin{figure}
  \centering
  \includegraphics[width=8cm]{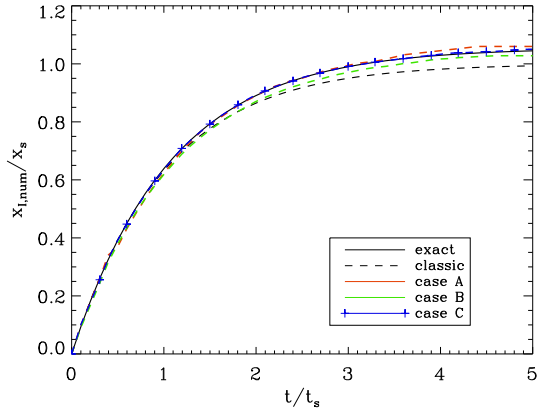}
  \caption{Time evolution of the ionization degree $\eta$ for the
  three test cases with different particle numbers and distributions
  red ($125$k particles), green ($100$k particles) and blue ($250$k
  particles) lines. 
  The black solid line denotes the exact solution, the black dashed
  line the classic (Str{\"o}mgren) solution. \label{stroem_eta_t}}
\end{figure}

In Fig. \ref{stroem_eta_t} the time evolution of the penetration
length $x_\mathrm{I}(t)$ is shown. The position of the front is calculated by
projection of the three-dimensional simulation along the y- and z-axis
onto the x-axis. 
Note that the analytical solution (cf
Eq. \ref{eq_x_vs_t}) is based on the 
idealized assumption that the medium is fully ionized ($\eta=1.0$). 
However, the precise solution of Eq. \ref{eq_delta_dt} in equilibrium
($d\eta/dt=0$) is
\begin{equation}
\label{x_stroem_exact}
x_\mathrm{s}\,\eta^2 = \frac{V_\mathrm{s}}{S} =\frac{F_\mathrm{Ly}}{\alpha_\mathrm{B}n^2}.
\end{equation}
In our simulations $x_\mathrm{s}\eta^2=1$pc is realized with $x_\mathrm{s}=1.05$pc and
$\eta = 0.976$. We call this the exact solution whereas the solution
assuming $\eta = 1.0$ will be referred to as classic solution.
Our simulations converge very well towards the exact solution.  Case A,
where the particles are initially placed on a grid, 
slightly overestimates the final value of $x_\mathrm{s}$,
while the low resolution simulation (case B) underestimates it.
Nevertheless, already with only $100$k this implementation shows a very
good agreement with the analytical curve. In the high
resolution simulation (case C) the numerical result lies right on top
of the predicted line.

\begin{figure}
  \centering
  \includegraphics[width=8cm]{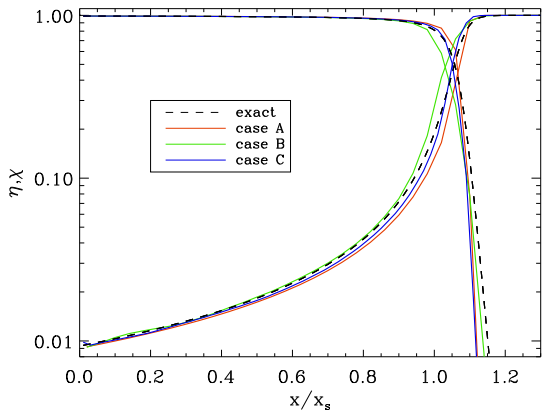}
  \caption{Ionization degree $\eta$ ($\approx 1$ at $x/x_\mathrm{s}=0$) and neutral gas fraction
  $\chi=1-\eta$ versus position for the  
  cases A (red), B (green) and C (blue).
  The dashed line represents
  the exact solution. \label{stroem_eta_x}}
\end{figure}
Fig. \ref{stroem_eta_x} shows the ionized fraction $\eta$ and the neutral
fraction $\chi=1-\eta$ after
$t=5t_\mathrm{s}$ at the end of the simulation. The numerical solution of
Eq. \ref{eq_eta_vs_x} is evaluated for the exact solution with a
penetration length of $x_\mathrm{s}=1.05$pc.
As expected from Fig. \ref{stroem_eta_t},
case A overestimates the front position, whereas
case B underestimates it. Again the high resolution
run C shows the best concordance and we can conclude that these results
fit well within the range of the code comparisons done by
\citet{2006MNRAS.371.1057I}. A more direct comparison to this work is
not possible due to the plane-parallel nature of the test performed
here. \footnote{Note 
  that in Fig. \ref{stroem_eta_x} the neutral fraction $\chi$
  converges towards a value of $10^{-2}$ at $x=0$pc in both the
  simulations and the exact solution whereas in
  \cite{2006MNRAS.371.1057I} $\chi$ is reaching much 
  lower values. This is due to the fact that in our simulations
  the irradiated surface stays constant whereas when simulating a
  point source this surface and thus $\chi$ can get infinitesimally
  small.} 

\subsubsection{Ionization by a time-varying source}
A more challenging test is the treatment of a time-varying source of
ionization. Although this situation is not very realistic for an O-star
it nevertheless provides a very good method to test the treatment of a
rapidly changing flux by the code.

To produce an ionization front that is traveling at a
    constant speed through a medium of
constant density it is sufficient to increase the flux per area
$F_\mathrm{Ly}$ linearly with time,
\begin{equation}
F_\mathrm{Ly}(t) = n v_\mathrm{f} +n^2\alpha_\mathrm{B}v_\mathrm{f}t,
\end{equation}
where $v_\mathrm{f}$ denotes the speed of the ionizing front. The first term on
the right hand side provides the ionization of the front, while the second term
compensates for the loss of flux due to recombinations on the way
towards the front. We assume a constant density of
$n = 10\textrm{cm}^{-3}$ and the velocity
of the front is set at $v_\mathrm{f} = 1.3\times 10^5 \textrm{cm s}^{-1}$. The other parameters are chosen as before.

\begin{figure}
  \centering
  \includegraphics[width=8cm]{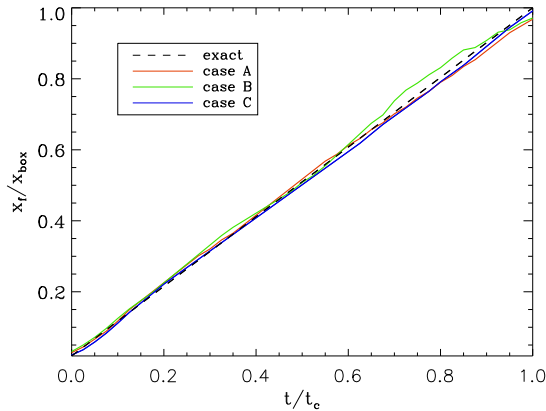}
  \caption{Numerical simulation of an ionization front that moves with constant speed through
  a medium of constant density. Plotted is the position of the front
  in units of the box length
  versus the time in units of the crossing time for the three cases A
  (red), B (green) and C (blue). \label{constslab}}
\end{figure}

Again the three initial conditions A, B and C from section \ref{stroem} are 
explored.
The results are shown in Fig. \ref{constslab}. As
before the simulations match the theoretical solutions closely.
In the beginning run A agrees very well with the solution. This
is due to the very low numerical noise in the Cartesian grid. However,
towards the end the low resolution leads to a deviation from the
analytic value.  
In case B
one can clearly see the effect of the noisy density distribution,
since for the recombination $R$ any error in the density leads to a
quadratic error in the absorption of the photons (cf
Eq. \ref{eq_rec}). Therefore, the position 
starts to oscillate around the exact position. This effect gets
stronger the further the front penetrates, as more material has to be
kept from recombining. Case C shows a very good
agreement with the exact solution, the resolution is high enough to
keep the noise in the density distribution low 
and thus the position of the front is followed precisely.

\subsubsection{Testing the refinement - Ionization by a constant
  source in a two-density medium}
\label{refine}
All tests up to now were independent of the implementation of
refinement, since in a constant density medium each particle occupies
roughly the same diameter $d_\mathrm{part}$ (see
Section \ref{implementation}). To verify the correct 
implementation of the
refinement we set up a simulation with a two-density medium. A lower
density gas phase with $n_\mathrm{1}=10\textrm{cm}^{-3}$ is set up in the left
half of the box and a
higher density medium with $n_\mathrm{2}=100\textrm{cm}^{-3}$ is placed at the
right half of the box. The density contrast is achieved via a
different number of SPH-particles in the different regions, the
particle masses are equal in the entire simulation. The required flux to ionize the simulation
domain up to a position $x_\mathrm{s}$ can be calculated by linear
superposition according to  Eq. \ref{x_stroem_exact} 
\begin{equation}
F_\mathrm{Ly}=\alpha_\mathrm{B}\,x_\mathrm{1}\,n_\mathrm{1}^2+\alpha_\mathrm{B}\,(x_\mathrm{s}-x_\mathrm{1})\,n_\mathrm{2}^2,
\end{equation}
where $x_\mathrm{1}=0.5$pc is the extent of the low-density region. The
simulation is set up with 550k randomly placed particles. The particle
noise is reduced for both regions separately as described in Section
\ref{stroem}. The 
penetration depth is set to $x_\mathrm{s}=1$pc. As soon as the equilibrium
state is reached the numerically calculated penetration lengths are
\begin{equation}
x_\mathrm{eq,unrefined} = 0.985\,x_\mathrm{s}
\,\,\,\,\textrm{ ; }\,\,\,\,
x_\mathrm{eq,refined} = 0.997\,x_\mathrm{s}
\end{equation}
for the unrefined and the refined case.
Very good agreement even with the unrefined code can be expected,
as we always use the SPH-density in the calculation which is independent of
the chosen bins for the calculation of the ionization and the recombination.

\begin{figure}
  \centering
  \includegraphics[width=8cm]{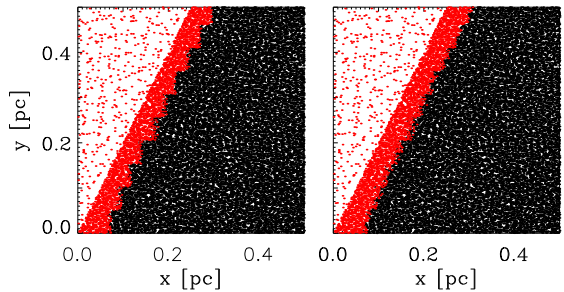}
  \caption{Effect of the refinement on a diagonal density
  step. Plotted are the SPH-particles in a $(0.5\textrm{pc})^3$ volume
  projected along the z-axis. Red: ionized particles ($\eta>0.1$), black: unionized
  particles. Left: without the inclusion of refinement. Right: with
  the inclusion of one level of refinement.  \label{refined}}
\end{figure}

Nevertheless, the refinement has an important geometric effect, which
becomes clear when assuming a density contrast with a discontinuity
which is not aligned vertical to the impinging radiation. We perform a
test with a diagonal density contrast between two regions with a
number density of $n_\textrm{low}=10\textrm{cm}^{-3}$ and
$n_\textrm{high}=200\textrm{cm}^{-3}$ respectively. Again the
particles are placed randomly and the noise is reduced (see Section
\ref{stroem}). A cubic domain of $(0.5\textrm{pc})^3$ including $25$k
particles is shown in Fig. \ref{refined}. In the unrefined case (left
hand side) the effect of the original bin-size of $\approx 0.05$pc can
be clearly seen as step-like features.  With refinement the density
contrast of $20$ leads to one level of refinement (since
$d_\textrm{part}/\Delta y\approx 1/2.7$, cf Eq. \ref{ref_crit}) and
the geometrical bias is already negligible (Fig. \ref{refined}, right
hand side). In the simulations in Section \ref{simulations} all five
levels of refinement lead to spatial resolution of the radiation in
our simulations as high as $10^{-3}$pc, therefore the radiation does
not produce any unphysical geometrical effects.

\subsection{Ionization with hydrodynamics}
\label{tests_hydro}
\subsubsection{Steady propagation of an ionizing front}
\label{lefloch}
This test was originally proposed by \citet{1994A&A...289..559L}. An
area of constant density is ionized by a photon flux which increases
linearly with time. This leads to a hydrodynamical shock wave
traveling at a 
constant speed. The number densities $n_\mathrm{i}, n_\mathrm{c}, n_\mathrm{0}$ in the ionized,
the compressed and the undisturbed medium can then be calculated from
the corresponding sound speeds $a_\mathrm{i}, a_\mathrm{c}, a_\mathrm{0}$. Let $u_\mathrm{i}$ be the speed
of the ionization front and  $u_\mathrm{s}$ be the speed of the shock front.
The jump condition for a D-type ionization front can be written
as
\begin{equation}
\frac{n_\mathrm{i}}{n_\mathrm{c}}=\frac{a_\mathrm{c}^2}{a_\mathrm{i}^2}=\frac{a_\mathrm{0}^2}{a_\mathrm{i}^2},
\end{equation}
since the compressed and the neutral medium have the same temperature
and thus the same sound-speed. At the isothermal shock the jump
conditions are
\begin{equation}
u_\mathrm{s}(u_\mathrm{s}-u_\mathrm{1})=a_\mathrm{0}^2 ,
\end{equation}
\begin{equation}
\frac{n_\mathrm{c}}{n_\mathrm{0}} = \frac{u_\mathrm{s}^2}{a_\mathrm{0}^2},
\end{equation}
where $u_\mathrm{1}$ is the gas velocity just inside the shock. With the
approximation of a thin shock it can be assumed that the ionization
front and
the shock front have the same speed $u_\mathrm{i} = u_\mathrm{s}$. For a detailed
derivation of the jump conditions see e.g. \citet{1991pagd.book.....S}.
Introducing the time derivative of the ionizing flux
$C=\mathrm{d}F/\mathrm{d}t$ the speed of the front can be calculated
similar to Eq. \ref{x_stroem}:
\begin{equation}
u_\mathrm{i} = \frac{C\alpha_\mathrm{B}}{n_\mathrm{i}^2} ( = u_\mathrm{s}).
\end{equation}
The jump conditions can then be rewritten to give the following relations:
\begin{equation}
n_\mathrm{c} = n_\mathrm{0}\frac{u_\mathrm{s}^2}{a_\mathrm{0}^2} = n_\mathrm{0}(\frac{C}{\alpha_\mathrm{B}n_\mathrm{i}^2a_\mathrm{0}})^2
\end{equation}
\begin{equation}
n_\mathrm{i} = n_\mathrm{c}\frac{a_\mathrm{0}^2}{a_\mathrm{i}^2} = (\frac{n_\mathrm{0}C^2}{\alpha_\mathrm{B}a_\mathrm{i}^2})^{\frac{1}{5}}
\end{equation}
\begin{equation}
u_\mathrm{1}=u_\mathrm{s}-\frac{a_\mathrm{0}^2}{u_\mathrm{s}}.
\end{equation}

To compare directly to previous results, we used the initial
conditions by \cite{1994A&A...289..559L}. The density is
$n_\mathrm{0}=100\textrm{cm}^{-3}$, the temperature is
$T_\mathrm{cold} = 100$K. The flux increases linearly with 
time at a constant rate of 
$\mathrm{d}F/\mathrm{d}t =
5.07\times10^{-8}\textrm{cm}^{-2}\textrm{s}^{-2}$, starting from
zero. 
As before the recombination parameter is set to $\alpha_\mathrm{B}=
2.7\times10^{-13}\textrm{cm}^3\textrm{s}^{-1}$ and the ionized temperature is
$T_\mathrm{hot}=10^4$K. Refinement is included.

The simulations are performed using the individual particle time-steps
of VINE. For the determination of the time-step we use the criteria
given in \cite{2008arXiv0802.4245W}. Here we will only review briefly
the parameters used. We use a combined time-step  
criterion based on the change in acceleration and velocity of the
particle with an accuracy parameter of $\tau_\mathrm{acc}=1.0$. In
addition, a CFL criterion is used with a tolerance
parameter of $\tau_\mathrm{CFL}=0.3$ and the modifications discussed
in Section \ref{timestepping}. We also use an additional time-step
criterion based on the maximum allowed change of
the smoothing length \citep[see][for details]{2008arXiv0802.4245W} with an
accuracy parameter of $\tau_\mathrm{h}=0.15$.
VINE employs a variable and time-dependent smoothing length, the
number of neighbours of each particle is on average
$n_\mathrm{neigh}=50$, but variations of $\pm20$ are allowed. 
The artificial viscosity of the SPH method is included in the standard way
\citep{1983MNRAS.204..715G} with the parameters $\alpha = 1$ and $\beta
= 2$ as implemented in VINE.

\begin{figure}
  \centering
  \includegraphics[width=8cm]{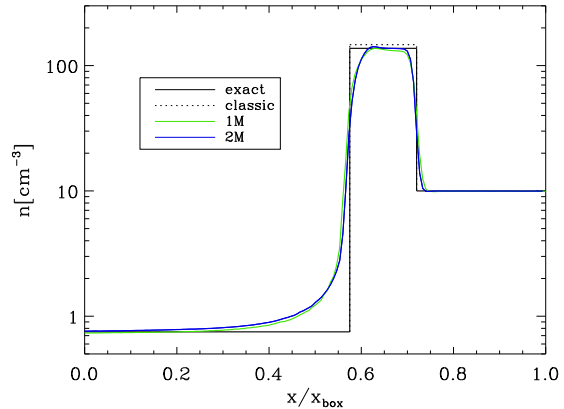}
  \caption{Number density versus position for the steady propagation of an
  ionizing front. The dashed line shows the classic solution, obtained
  by using a value of $T_\mathrm{hot,a}=10^4$K for the hot gas. The solid line
  corresponds to the analytic solution for a more realistic value of
  $T_\mathrm{hot,r}=9200$K for the hot gas. Blue and green lines show the simulations
  at different resolutions.\label{leflochfig}} 
\end{figure}

We performed simulations with 1 and 2 million particles in a cubic
simulation domain. The particles are distributed randomly and then
left to relax according to Section \ref{stroem}. The higher resolution
compared to the test in 
Section \ref{tests_no_hydro} is necessary to follow the hydrodynamical
shock precisely.
As in the tests before, assuming a fully ionized gas with $\eta =1.0$ and
thereby using a value of $T_\mathrm{hot,a}=10^4$K for the calculation of the
sound-speed of the hot gas, does not correspond to the simulations (see
equation \ref{x_stroem_exact}). Instead a much better agreement can be
achieved when the real temperature of the gas in the simulations,
$T_\mathrm{hot,r}=9200$K, is used (since $\eta = 0.92$ on average in the
ionized region). For this 
more realistic temperature the simulations are in very good agreement 
and converge towards the analytic solution with increasing
resolution (see Fig. \ref{leflochfig}). This can
also be seen in Table \ref{leflochtbl}.
\begin{table}
\begin{center}
\begin{tabular}{cccccc}
\hline
& Classic & Exact & Grid & SPHI & iVINE  \\
\hline
$n_\mathrm{c}\,(\textrm{cm}^{-3})$ & 147 &  137 &169 & 155 & 138 $\pm$ 6\\
$n_\mathrm{i}\,(\textrm{cm}^{-3})$ & 0.734 & 0.747 & 0.748 & 0.75 & 0.743 $\pm$ 0.01\\
$u_\mathrm{i}\,(\textrm{km s}^{-1})$ & 3.48 & 3.37&3.36 & 3.43 & 3.34 $\pm$ 0.18 \\
$u_\mathrm{1}\,(\textrm{km s}^{-1})$ & 3.24 & 3.12 & - & - & 3.13 $\pm$ 0.04\\
\hline
\end{tabular}
\caption{Comparison of analytical and numerical results for the test
  including hydrodynamics and ionization. The iVINE data is obtained
  from the 2 million particle run, the errors given are 1$\sigma$. The
  grid data is taken from \citet{1994A&A...289..559L}, the SPHI data
  from \citet{2000MNRAS.315..713K}. The analytical values differ from
  the previous work due to a higher accuracy in our
  calculations.\label{leflochtbl}}
\end{center}
\end{table}

\subsection{Performance}
\label{perform}
To test the performance of the parallel iVINE code with increasing
number of processors we choose the simulation described in
Section \ref{simulations} at a later stage
and compute one time-step on different numbers of CPUs.
The parallel scaling of the various parts of the underlying VINE code is
discussed in detail in \cite{2008arXiv0802.4253N}. For our test, we
use a SGI Altix 3700 Bx2 
supercomputer.  In total, the ionization uses only a few percent of
the total computational time. The precise values range from 2.32\% on
2 CPUs to 2.70\% on 16 CPUs and 2.86\% on 32 CPUs. 

When refinement is used, these values change to 8.52\% on 2
CPUs and 8.73\% on 32 CPUs. Although the ionization takes up 
relatively more time in this case, the difference in the 
calculation time between the number of CPUs gets smaller. This is to
be expected, as the refinement is calculated inside the bins and this
part of the implementation is parallelized very efficiently (each bin
is independent of the other bins).

This test shows that the additional  cost of our implementation of
ionizing radiation in SPH is always much smaller than the cost for
other implemented physics, like gravity and hydrodynamics. In
particular, our new ray-tracing-scheme shows a substantial speedup
compared to the algorithm by \cite{2000MNRAS.315..713K}, where the
path-finding alone took up about 50\% of the total computational time
\citep{1999PhDT........18K}. Another drawback of their approach is
that for every particle the optical depth is calculated along a path
towards the source until a particle closer to the source with an
already calculated optical depth is found. This is a highly serial
approach and thus the scheme of \cite{2000MNRAS.315..713K} does not
lend itself easily to an efficient parallelization.

\section{Radiation driven implosion}
\label{simulations}
As a first application of iVINE we 
model the radiation
driven implosion of an otherwise stable molecular cloud core. This
approach is very similar to \citet{2003MNRAS.338..545K} 
 but at ten times higher mass resolution. 
A marginally stable Bonnor-Ebert sphere (BES) \citep{1956MNRAS.116..351B}
with a radial pressure profile defined by 
\begin{equation}
  \frac{1}{r^2}\frac{d}{dr}(\frac{r^2}{\rho}\frac{dp}{dr})=-4\pi G\rho
\end{equation}
is exposed to UV-radiation from a nearby source.
The temperature of the sphere is $T=10$K, the peak
density is $n_\mathrm{max} = 10^3\textrm{cm}^{-3}$,
and the gas is initially at rest (i.e. no turbulence). 
The total mass contained in
the sphere is $96 \textrm{M}_\odot$ and the radius is $1.6$pc. We
embed the sphere into cold gas ($10$K) with a constant density
corresponding to the cutoff-density at the edge of the sphere.
These simulations where performed
with $2.2\times10^6$ particles resulting in a particle mass of
$7.2\times10^{-5}\textrm{M}_{\odot}$. Self gravity is included. 
The cooling timescale ($t_\mathrm{cool} < 0.3$kyr) is much shorter than
any other timescale involved in our simulations (e.g. the crossing
timescale of the hot gas is $t_\mathrm{hot} \approx
70$kyr). Thus, we treat the non-ionized gas with an isothermal equation of
state ($\gamma=1$). The ionized gas is assigned a temperature
according to Eq. \ref{eq_temperature} with $T_\mathrm{hot}=10^4$K and
$T_\mathrm{cold} =10$K and then treated isothermally as well.

The artificial
viscosity and the criteria for the individual time-steps are the same
as in Section \ref{lefloch} ($\alpha = 1$, $\beta = 2$, $\tau_\mathrm{acc}=1.$,
$\tau_\mathrm{CFL}=0.3$ and $\tau_\mathrm{h}=0.15$). In addition, we
use a multipole acceptance  criterion (MAC) for the tree based
calculation of gravitational forces according to
\cite{2001NewA....6...79S} as implemented by
\cite{2008arXiv0802.4245W} with a tree accuracy parameter of
$\theta=5\times10^{-4}$.
The implementation of the SPH smoothing kernel and the gravitational
softening length in VINE are equal at all times. The number of neighbours is
set to $n_\mathrm{neigh}=50\pm20$.
The hydrodynamical boundaries are periodic
in the y- and z- direction, and open in the x-direction. This
resembles the situation around a massive O-star 
where the material is allowed to move freely in the radial direction
while at the sides similar material is existing. Gravitational forces are
calculated by just taking into account the self-gravity of the gas and
no external or boundary effects. 
This is reasonable as the total simulation time ($<600$kyr) is much
shorter than the free-fall time ($t_\mathrm{ff}\approx 1.5$Myr).
In the simulations the radiation is impinging from the left hand
 side. We perform three different simulations, differentiated by the 
 penetration length in the surrounding medium relative to the box size $C=x_\mathrm{s}/x_\mathrm{box}$:
\begin{itemize}
\item Simulation HF (high flux):
\item [  ]$F_\mathrm{0} = 9.0\times 10^9\textrm{photons cm}^{-2}\textrm{s}^{-1} => C\approx 1.0$
\item Simulation IF (intermediate flux): 
\item [  ]$F_\mathrm{0} = 4.5\times 10^9\textrm{photons cm}^{-2}\textrm{s}^{-1} =>
  C\approx 0.5$
\item Simulation LF (low flux):
\item [  ]$F_\mathrm{0} = 9.0\times 10^8\textrm{photons cm}^{-2}\textrm{s}^{-1} =>
  C\approx 0.1$
\end{itemize}
This corresponds to the
 molecular cloud being placed inside (HF), at the border (IF) and
 outside (LF) of the Str{\"o}mgren sphere. 
 The evolution of the BES for all three cases is shown in
 Fig. \ref{sim_BE}.

\subsection{Dynamical evolution}
The general evolution of a simulation of this kind is as follows:
As soon as the simulation starts, a R-type ionization front is driven
into the medium. As it can be expected from Section \ref{stroem}, the front
reaches the Str{\"o}mgren radius $x_\mathrm{s}$ of the diffuse gas within a
few recombination timescales ($5t_\mathrm{rec} \approx5$kyr).
After a sound crossing timescale ($t_\mathrm{hot} \approx 70$kyr) the
hot gas reacts to its change in pressure and starts to drive a
shock front into the cold gas - a D-type front evolves. This front
will affect the morphology of the BES. In the
following we describe the individual cases in more detail.

\subsubsection{Simulation HF (high flux)}
Due to the high flux (see Fig. \ref{sim_BE} first column), the R-type
front is able to propagate very far into the simulation domain. A
bow-like shock structure around the edge of the BES evolves ($t\approx
100$kyr).
The shock front running into the denser parts of the cloud is
slowed down, so that the front starts to "wrap around" the cloud.
Soon the two flanks are approaching each other while the center of the
shock is held back by the dense innermost region (Fig. \ref{sim_BE}
third row, first column, $t\approx 100 $kyr). As the two sides
finally  collide an elongated  filament forms which is gravitationally
unstable. In Fig. \ref{sim_BE_final} we show this filament at the
final stage of our simulations. In comparison runs without
self-gravity the two shock fronts cross each other and the 
cloud disperses. With 
the inclusion of self-gravity the filament
becomes gravitational unstable and is triggered into collapse. In
fact the core fragments into several objects, as will be discussed in
a subsequent paper. The resolution limit according to
\citet{1997MNRAS.288.1060B} is
$n_\textrm{max}=2\times10^{10}\textrm{cm}^{-3}$ in these
simulations. As soon as this limit is reached the local Jeans 
mass is smaller than the mass of 100 particles and artifical 
fragmentation can occur. Thus, we stop the simulations at this point.

\subsubsection{Simulation IF (intermediate flux)}
With an intermediate flux the R-type front penetrates
much less into the gaseous medium (see Fig. \ref{sim_BE} second
column). Thus, the  front does not wrap around the 
sphere as much as in Simulation HF. 
As soon as the hot gas reacts to its increased temperature a flattened
shock with a much smaller curvature than in Simulation HF forms. In
addition, the motion in the hot gas forces the flanks on both sides of
the main shock inwards, which can be seen in the velocities of the hot
gas in Fig. \ref{sim_BE} (third row, second column, $t\approx
100 $kyr). These motions are due to the periodic boundaries on the
upper and lower edge. Otherwise the gas could stream away
freely. However these boundaries are justified by the fact that 
the molecular cloud is completely surrounding the O-star.
In the further evolution the flanks approach each other similar to
Simulation HF and the central region becomes unstable and fragments
(see Fig. \ref{sim_BE_final}).

\subsubsection{Simulation LF (low flux)}
\begin{figure*}
  \centering{
  \includegraphics[width=16.8cm]{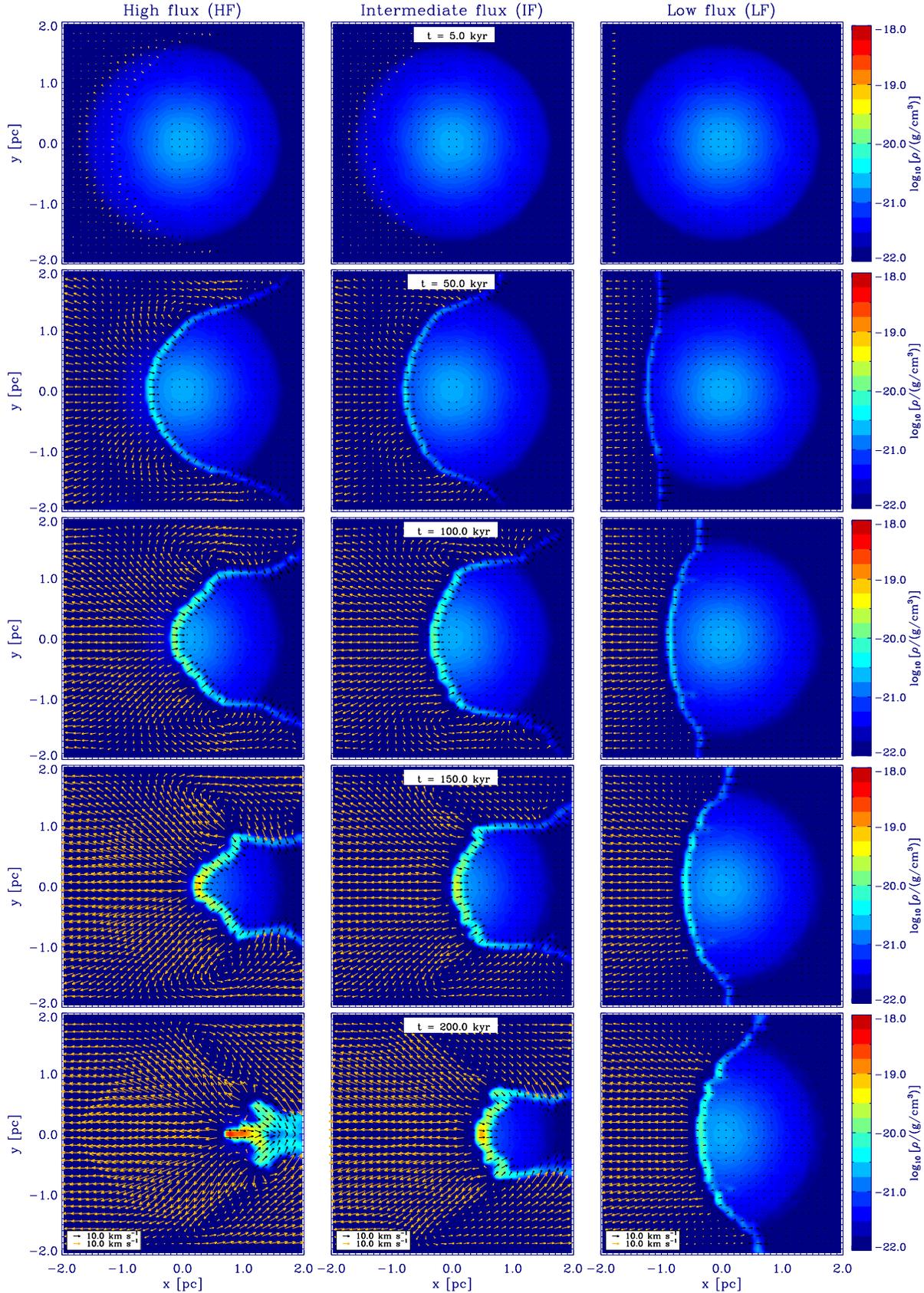}
  \caption{
Time evolution of the driven collapse of a Bonnor-Ebert sphere ionized
  by a source with high flux (first column), intermediate flux (second
  column) and low flux (third column). The simulation volume is a cube
  with sides $4$pc long, the UV-radiation is impinging
  from the left hand side. Color coded is the density of the central
  cold gas slab. Yellow arrows denote the velocities of  the hot 
  gas, black arrows the motion of the cold gas. Density and velocities
  are averaged across a slice of $0.125$pc in the z-direction. Each row
  shows the three simulations at a different time. 
}\label{sim_BE}}
\end{figure*}

The very low flux in this case only leads to a R-front which barely
reaches the sphere (see Fig. \ref{sim_BE} third
column). Therefore, the D-front starts as a nearly
plane-parallel shock wave in front of the BES.
This shock sweeps away much more material than in the high and intermediate
flux cases, where the material is concentrated in the
center. Nevertheless, as the shock propagates further the very center
of the sphere gets compressed and becomes gravitationally unstable.
In contrast to Simulations HF and IF there is no sign of fragmentation in
the unstable region.

\subsection{Structure, collapse timescales and final mass assembled}
\label{threecore}
\begin{figure}
\centering{
\includegraphics[width=8.cm]{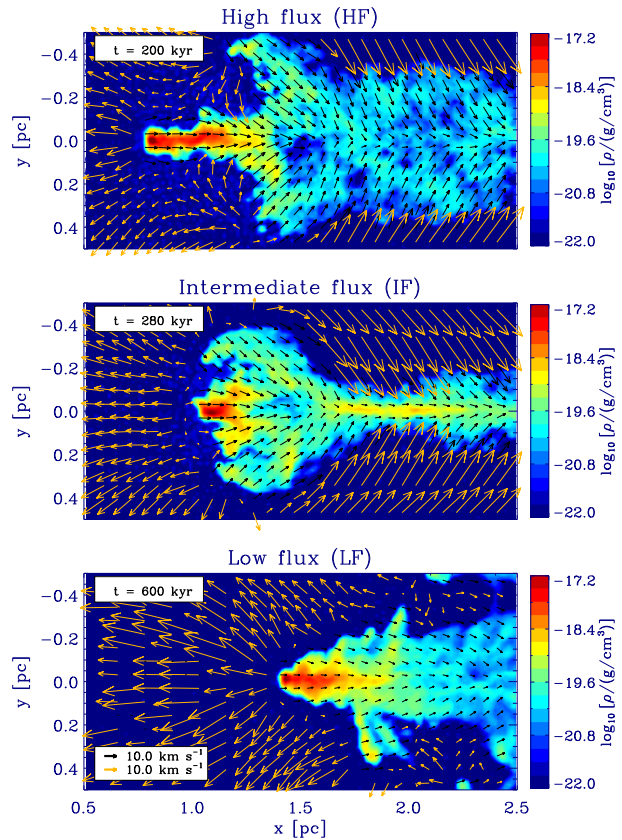}
  \caption{Final stage of the three simulations. Color coded is
  the density in the central slab. Yellow arrows denote the velocities of
  the hot gas, black arrows the motion of the cold gas. Density and
  velocities are averaged across a slice of $0.0625$pc in the
  z-direction. The time of 
  the collapse as well as the displacement of the fragment clearly
  depend on the initial flux. Furthermore the velocity of the cold gas (black
  arrows) is decreasing with decreasing flux. The core always forms at
  the very tip of the filament.
}\label{sim_BE_final}}
\end{figure}

A close look at the final structure of the collapsing filaments of the
three simulations (Fig. \ref{sim_BE_final}) reveals that in all three
simulations the core forms at the tip of an elongated filament, which
might be eroded in the future. This matches exactly the observed head to tail
structure described in Section \ref{intro}. The core regions have an
extent of just $0.02-0.05$pc, which corresponds very closely to
e.g. the findings of \cite{2001A&A...365..440M} in the the Perseus
star cluster. They observe 8 Class 0 protostars
with compact envelopes
($R_\mathrm{out}<10^4\textrm{AU}\approx0.05$pc). In 
addition they are denser by a factor of 3-12 than it would be expected
from the standard collapse model, which would suggest densities of
$n\approx10^6\textrm{cm}^{-3}$ (see e.g. Walch et al. 2008, in
preparation). \cite{2001A&A...365..440M} suggest that this higher
densities are due to external disturbances initiating the collapse,
which agrees very well with our simulations.
Following the observations we define a core as all material with a
density higher than $n_\mathrm{crit} = 10^7\textrm{cm}^{-3}$ in a region of
$R_\mathrm{crit}=0.02$pc (which is roughly a Jeans length at a density of
$n_\mathrm{crit}$) around the peak density.

\begin{figure}
\vspace{.4cm}
  \centering{
  \includegraphics[width=8cm]{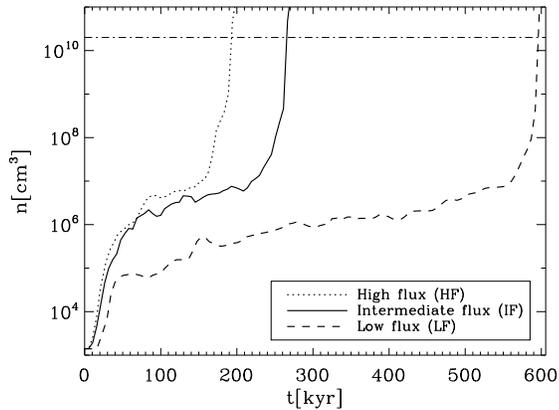}
  \caption{The maximum number density versus time for the three different
  simulations. In the higher flux cases HF and
  IF the collapse happens much earlier than in the low flux case LF. 
The dash-dotted line represents the resolution
  limit as given by \citet{1997MNRAS.288.1060B}. \label{rhomax}}}
\end{figure}

We plot the evolution of the maximum density in Fig. \ref{rhomax}. 
In all three simulations after the first phase of compression by the
hot gas a meta-stable phase at densities between
$10^6\textrm{cm}^{-3}$ and $10^7 \textrm{cm}^{-3}$ 
can be seen. This fits very nicely to the structure of observed cores
described above. The duration of this phase depends on the
initial flux (HF: $90$kyr, IF: $155$kyr, LF: $290$kyr). 
In addition, we find evidence that the filaments collapse earlier in
cases with a higher flux. The collapse happens  at $t=200$kyr,
$t=280$kyr and $t=600$kyr in Simulation HF, IF and LF, respectively.
Observations of triggered star formation tend to show the
same trend (see \citealt{2005ApJ...624..808L},
\citealt{2008AJ....135.2323I}) - the younger the star,
the further it is away from the ionizing source. This can not be
explained by just attributing it to the speed of the R-type front. As
seen in Section \ref{stroem} the 
crossing time for the R-type front is of the order of a few kyr, whereas
any observed age-spread is of the order of several hundred kyr. To
explain this huge 
spread the position of the density enhancement relative
to the Str{\"o}mgren radius has to be considered. As we show
decreasing the flux and thereby increasing
the distance to the source can delay collapse and star formation 
by $0.08-0.4$Myr.

In IC 1396N \citet{2007ApJ...654..316G} report a T Tauri (Class
II and Class III stars) population with ages $\approx0.5-1$
Myr. In addition, $0.3-0.5$pc further away from the ionizing source HD 206267
(an O6.5f-type star), there is an embedded population of Class 0/I
protostars with ages $\approx 0.1$Myr. This can be compared to our simulations
where e.g Simulation IF represents gas clumps closer to the source
which start to form stars $0.3$Myr earlier than Simulation LF. So at
the time the embedded stars start to form in Simulation LF the stars
of Simulation IF would be no longer embedded and represent the Class
II/III stars population. In fact in our simulations the spread of a
few hundred kyr is smaller than in the observations. This difference
could be attributed to the classification of the protostars as
discussed below.

Besides the age-spread one can look at the velocities of the front and
core. From
Fig. \ref{sim_BE} it can be seen that the shock front travels with a
speed of $3-7$km/s, depending on the initial flux. Most observational
estimates provide a front speed $<1\textrm{km/s}$
\citep{2004A&A...414.1017T}, leading to a difference of almost an order of 
magnitude between observations and simulations which has been noted before
(see e.g. \citealt{2007ApJ...654..316G}). The age estimates of the
YSOs are mainly based on their classification by the spectral energy
distribution (SED). A Class 0/I object is deeply embedded, therefore
the short micrometer wavelengths are much weaker due to absorptions in
the envelope when compared to Class II/III objects. This allows for a
clear distinction between both types e.g. in the IRAC $[3.6]-[4.5]$
versus $[5.8]-[8]$ color diagram as shown by
\cite{2005ApJ...629..881H}. We suggest that in the case of triggered star 
formation, the ionizing radiation could strip the envelope of a
YSO, thereby unveiling the central object in short micrometer
wavelengths. Thus, the observed Class II/III SED could be caused by a
much younger Class 0/I protostar with a removed envelope. This would
reduce the estimated age spread, thereby increasing the estimated
speed of the shock front and  finally leading observations and
simulations to agree. This assumption will be subject to further examination.

A dependence on distance is also seen for the velocities of the cold
filaments (see the velocities of the cold gas (black arrows) in
Fig. \ref{sim_BE_final}). 
The precise velocities of the cores in the Simulations HF, IF and LF are
$8.4\textrm{km/s}$, $7.6\textrm{km/s}$ and $5.1\textrm{km/s}$,
respectively. Again, the closer the core is to the OB-association the
higher is its velocity. Although this small differences are not
observable yet it is worth noticing that the cores themselves have  bulk
velocities which are slightly higher (by about $0.5-1$km/s) than the
rest of the filament. However, this effect may get weaker as the core
gets slowed down while sweeping up the rest of the filament. 

The final mass assembled does not show a dependence on the initial
distance. In Simulation HF 
the core consists of $6.0\textrm{M}_{\odot}$ in Simulation IF of
$7.4\textrm{M}_{\odot}$ and in Simulation LF of $2.8\textrm{M}_{\odot}$.
The filaments as a whole have masses of $61.5\textrm{M}_{\odot}$, $75.3\textrm{M}_{\odot}$ or
$67.4\textrm{M}_{\odot}$, respectively.
It is obvious
that the most effective scenario is 
Simulation IF. Here, 
the ionization encompasses most of the sphere and thus the shock front
is not nearly as plane-parallel as in Simulation LF and does not sweep
away as much material. On the other hand, less material gets evaporated
by the ionization since the flux is lower than in Simulation HF. 
Overall the final masses of the collapsing cores fit the
observations well. Assuming a star formation efficiency of 30\% (see
e.g. \citealt{2008ApJ...672..410L}), we find masses from
$0.84\textrm{M}_{\odot}$ to $2.2\textrm{M}_{\odot}$ which 
agrees with the observed range from classical T Tauri up to Herbig Ae/Be
stars (see e.g \citealt{2007ApJ...657..884L},
\citealt{2007arXiv0711.1515S}).

\section{Summary and Discussion}
\label{summary}
We present iVINE, a new implementation of UV-radiation into the tree-SPH
code VINE. It uses a plane-parallel geometry which renders the code
most suitable to perform high resolution studies of the small scale
effects of e.g. ionization and turbulence in the surrounding of young
massive stars.  It is efficiently parallelized and very
fast, as only 2\%-8\% of the total computational time are used for the
calculation of ionization. 
The comparison with analytic solutions shows that iVINE treats
time-dependent ionization as well as the resulting heating effects
precisely and convergently.

We base our numerical implementation of ionizing radiation on several
assumptions. 
First, we use a simplified prescription for the radiative transfer by
e.g assuming a monochromatic flux. 
Second, we neglect UV absorption by dust, which would lower the total
UV flux. 
Third, we do not include a full treatment of recombination zones.
In our simulations the ionized gas which gets shaded is assumed to
recombine immediately. In addition, the gas in the shaded regions does
not get heated by irradiation from the hot gas surrounding it. These
effects require a precise time-dependent treatment of heating and
cooling processes by ionization and recombination as well as a
treatment for the scattering of photons. An implementation of this
effects is planned in a future version of the code.

As an application we investigate radiation driven implosion of a
marginally stable Bonnor-Ebert sphere. 
We show that these spheres are indeed driven into gravitational
collapse. The resulting cores are in the observed mass range. They
are more compact and a factor of $\approx 10$ more dense than
it would be expected in a more quiescent environment. This fact fits
very well with the observations of star formation in a clustered
environment.  By comparing simulations with three different UV-fluxes 
we show that there is a clear dependence of the final mass and the age
of the collapsed core on the position of the preexisting density
enhancement relative to the Str{\"o}mgren radius. 
Our findings that the onset of star formation is delayed by
$0.08-0.4$Myr, depending on the position, are in good agreement with
observations of the age spread in bright rimmed clouds.
The velocity of the triggering shock is an order of magnitude higher
than the observational estimates. This discrepancy has been noted
before. We suggest that this can be 
attributed to the ionizing radiation stripping the envelope from a
Class 0/I star. Thereby it might be classified as an Class II/III star,
leading to an higher age-spread between the observed
objects. Correcting for this effect would increase the estimated velocity
of the shock front and thus lead simulations and observations towards
agreement.

\section*{Acknowledgments}
We would like to thank the referee, James Dale, for his valuable
comments on the manuscript.

This research was supported by the Deutsche Forschungsgesellschaft (DFG),
SFB 375 and by the DFG cluster of excellence "Origin and
Structure of the Universe".

All simulations were performed on a SGI Altix 3700 Bx2 supercomputer that was
partly funded by the DFG cluster of excellence "Origin and Structure of the Universe".

\bibliographystyle{mn2e}
\bibliography{references}
\label{lastpage}

\end{document}